\documentstyle[twocolumn,prb,aps,epsfig,amsfonts]{revtex}

\begin{document}
\draft
\preprint{}

\newcommand{\1}{{\bf \scriptstyle 1}\!\!{1}}
\newcommand{\I}{{\rm i}}
\newcommand{\p}{\partial}
\newcommand{\D}{^{\dagger}}
\newcommand{\bx}{{\bf x}}
\newcommand{\bk}{{\bf k}}
\newcommand{\bv}{{\bf v}}
\newcommand{\bp}{{\bf p}}
\newcommand{\bu}{{\bf u}}
\newcommand{\bA}{{\bf A}}
\newcommand{\bB}{{\bf B}}
\newcommand{\bK}{{\bf K}}
\newcommand{\bL}{{\bf L}}
\newcommand{\bP}{{\bf P}}
\newcommand{\bQ}{{\bf Q}}
\newcommand{\bS}{{\bf S}}
\newcommand{\bH}{{\bf H}}
\newcommand{\balpha}{\mbox{\boldmath $\alpha$}}
\newcommand{\bsigma}{\mbox{\boldmath $\sigma$}}
\newcommand{\bSigma}{\mbox{\boldmath $\Sigma$}}
\newcommand{\bOmega}{\mbox{\boldmath $\Omega$}}
\newcommand{\bpi}{\mbox{\boldmath $\pi$}}
\newcommand{\bphi}{\mbox{\boldmath $\phi$}}
\newcommand{\bnabla}{\mbox{\boldmath $\nabla$}}
\newcommand{\bmu}{\mbox{\boldmath $\mu$}}
\newcommand{\bepsilon}{\mbox{\boldmath $\epsilon$}}

\newcommand{\iLambda}{{\it \Lambda}}
\newcommand{\cA}{{\cal A}}
\newcommand{\cD}{{\cal D}}
\newcommand{\cL}{{\cal L}}
\newcommand{\cH}{{\cal H}}
\newcommand{\cI}{{\cal I}}
\newcommand{\cO}{{\cal O}}
\newcommand{\cR}{{\cal R}}
\newcommand{\cU}{{\cal U}}
\newcommand{\cT}{{\cal T}}

\newcommand{\be}{\begin{equation}}
\newcommand{\ee}{\end{equation}}
\newcommand{\bea}{\begin{eqnarray}}
\newcommand{\eea}{\end{eqnarray}}
\newcommand{\beqa}{\begin{eqnarray*}}
\newcommand{\eeqa}{\end{eqnarray*}}
\newcommand{\nn}{\nonumber}
\newcommand{\DD}{\displaystyle}

\newcommand{\ba}{\left[\begin{array}{c}}
\newcommand{\baa}{\left[\begin{array}{cc}}
\newcommand{\baaa}{\left[\begin{array}{ccc}}
\newcommand{\baaaa}{\left[\begin{array}{cccc}}
\newcommand{\ea}{\end{array}\right]}

\twocolumn[
\hsize\textwidth\columnwidth\hsize\csname
@twocolumnfalse\endcsname

\title{Spin tunneling and topological selection rules for integer spins
}

\author{Michael N.~Leuenberger\cite{email1} and Daniel
Loss\cite{email2}}
\address{Department of Physics and Astronomy, University of Basel \\
Klingelbergstrasse 82, 4056 Basel, Switzerland}

%\date{\today}
\maketitle

\begin{abstract}
We present topological interference effects for the
tunneling of a single large spin, which are
caused by the symmetry of a general class of
magnetic
anisotropies. The interference originates from spin Berry
phases associated with different tunneling paths exposed to  the same
dynamics.  Introducing a generalized path integral for coherent spin
states, we evaluate transition amplitudes between ground as well as
low-lying excited states. We show that these interference effects lead
to topological selection rules and spin-parity effects
for integer spins that agree
with quantum selection rules and which thus provide a generalization of
the Kramers degeneracy to integer spins. Our results apply to the 
molecular magnets Mn$_{12}$ and Fe$_8$.
\end{abstract}

]
\narrowtext

In recent years the  tunneling of spin has attracted much
attention since experiments provided strong evidence for the existence
of spin tunneling in molecular magnets, such as Mn$_{12}$-acetate
(Mn$_{12}$)\cite{Paulsen et al,Friedman,Thomas} and
Fe$_8$-triazacyclononane (Fe$_8$).\cite{Sangregorio,Wernsdorfer}
In Refs.~\onlinecite{Paulsen et al,Friedman,Thomas,Sangregorio} the relaxation
time $\tau(H_z)$ of the magnetization as a function of the
longitudinal magnetic field $H_z$ was measured, with $\tau(H_z)$
exhibiting
large  peaks due to thermal-assisted tunneling of the
spin.\cite{Fort,LLMn12} In Ref.~\onlinecite{Wernsdorfer}, the incoherent Zener
rate\cite{LLFe8} as
a function of the transverse magnetic field $H_x$ was measured, from
which quantum
oscillations of the tunnel splitting $E_{mm'}(H_x)$ can be
observed.\cite{Wernsdorfer}
In a coherent spin-state path-integral approach,\cite{Klauder} these
quantum
oscillations and their associated spin-parity effects in the case of
quadratic anisotropies can be viewed as a result of interfering Berry
phases carried by spin tunneling paths of opposite
windings,\cite{Loss,Delft} which are modified in the presence of a
transverse-field $H_x$ (Ref.~\onlinecite{Garg}) or crystal fields of cubic symmetry
with multiple degenerate minima.\cite{Kalatsky}

The results obtained in the present work can be summarized as follows.
If the
Hamiltonian for a single spin
$s\gg 1$ contains a magnetic anisotropy
of
arbitrary order $n$ that is transverse to the easy axis
and symmetric under rotations of $2\pi/n$ around the easy axis, there
are
$n$ different paths that possess the same dynamical action but
different
spin Berry phases and which thus interfere  constructively or
destructively. Using a spin path integral that we generalize to excited
states
we will show that the
tunneling of the spin between ground
or low-lying
excited states that are doubly degenerate,
$\left|m\right>$ and $\left|-m\right>$, $|m|\leq s$,
is completely
suppressed if $n$ is not a
divisor of $2m$ (i.e., $2m/n\not\in{\Bbb Z}$); this result represents
a generalization of the
Kramers degeneracy to integer spins $s$.
We also provide the connection to quantum selection rules using standard
spin
algebra, and
we discuss the effect
of magnetic fields on these selection rules.

We consider a general single-spin
Hamiltonian
$\cH_{z,n}=-AS_z^2+B_n(S_+^{n}+S_-^{n})$,
with easy-axis
($A$) and transverse ($B_n$) anisotropy
constants satisfying
$A\gg B_n>0$.
Here, $n$ is an even integer, i.e., $n=2,4,6,\ldots$, so that $\cH_z$ is
invariant
under time reversal. Such Hamiltonians are relevant for molecular
magnets such as Mn$_{12}$ (Refs.~\onlinecite{Fort} and \cite{LLMn12}) and
Fe$_8$.\cite{Wernsdorfer,LLFe8}
The corresponding classical anisotropy
energy
$E_{z,n}(\theta,\phi)=-As^2\cos^2\theta+B_ns^n\sin^n\theta\cos(n\phi)$
has the shape of a double-well potential, with the easy axis pointing
along the $z$ direction. It is obvious that the anisotropy
energy remains invariant under rotations around the $z$ axis by multiples
of the angle $\eta=2\pi/n$, i.e.,
$\cR_{z,\eta}E_z(\theta,\phi)=E_z(\theta,\phi+\eta)=E_z(\theta,\phi)$.
It is convenient to describe the symmetry of the anisotropy energy by
the cyclic group
$C_{\alpha,n}=\{\cR_{\alpha,\eta},\cR_{\alpha,2\eta},\ldots,\cR_{\alpha,(n-1
)\eta},\cR_{\alpha,n\eta}=\cI\}$,
where $\cR_{\alpha,\varphi}$ performs rotations about the $\alpha$-axis,
$\alpha=x,y,z$, by the angle $\varphi$, and $\cI$ denotes the identity
operator.

For the following calculation it
proves favorable to choose the easy axis along the $y$ direction. Then the
spin Hamiltonian and the anisotropy energy are changed into
\be
\cH_{y,n}=-AS_y^2+B_n(S_+^{n}+S_-^{n}),
\label{hamiltonian_y}
\ee
where now $S_\pm=S_z\pm iS_x$,
and
\bea
E_{y,n}(\theta,\phi) & = &
-As^2\sin^2\theta\sin^2\phi+B_ns^n[(\cos\theta \nn\\
& & +i\sin\theta\cos\phi)^n+(\cos\theta-i\sin\theta\cos\phi)^n].
\label{energy_y}
\eea
 We note again that
$C_{y,n}E_{y,n}=E_{y,n}$, which will play a central role in what
follows.

We are interested in the tunneling between the eigenstates
$\left|m\right>$ and
$\left|-m\right>$ of $-AS_y^2$, corresponding to the global minimum
points
$(\theta=\pi/2,\phi=-\pi/2)$ and $(\theta=\pi/2,\phi=+\pi/2)$ of
$E_{y,n}$. For this we evaluate the imaginary time transition amplitude
between these points. For the ground-state tunneling ($m=s\gg 1$) this
can be done by means of the coherent spin-state path
integral\cite{Loss,Fradkin}
\be
\left<-\frac{\pi}{2}\left|e^{-\beta\cH}\right|+\frac{\pi}{2}\right>=
\int_{\pi/2,-\pi/2}^{\pi/2,+\pi/2}\cD\Omega\; e^{-S_{\rm E}},
\label{amplitude}
\ee
where $\beta=1/k_BT$ is the inverse temperature,
$\cD\Omega=\Pi_\tau d\Omega_\tau$, $d\Omega_\tau=[4\pi/(2s+1)] d(\cos\theta_\tau) d\phi_\tau$ the Haar measure
of the $S^2$ sphere, and $S_{\rm E}=\int_0^\beta d\tau[is\dot{\phi}(1-\cos\theta)+E_{y,n}]$ the Euclidean action,
where the first term in $S_{\rm E}$ defines the Wess-Zumino (or Berry
phase) term,\cite{Fradkin} which gives rise to topological
interference effects for spin tunneling.\cite{Loss,Delft,Garg,Kalatsky}

\begin{figure}[tb]
  \begin{center}
    \leavevmode
\epsfxsize=8cm
\epsffile{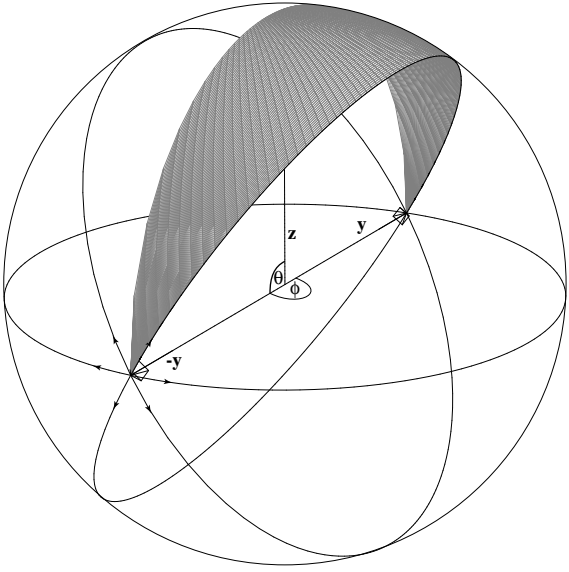}
  \end{center}
\caption{Interference between $n=6$ paths (small arrows) moving from
${\bf -y}$ to ${\bf y}$ direction (large arrows), which is destructive
if $n$ is not a divisor of $2m$. The grey surface area $\Phi(m)/m$
depicts the
Berry phase $\Phi(m)$ of the upper right path. Note that there is no simple
pairwise interference; instead, all six paths interfere with each other
[see Eq.~(\protect\ref{phases})].}
\label{interference2}
\end{figure}

For the tunneling between low-lying excited states ($1\ll |m|\lesssim
s$)
we now derive a generalized coherent spin-state path integral. It is
known\cite{Perelemov} that one can construct a coherent state system
$\{T\left|\psi_0\right>\}$ out of any starting vector
$\left|\psi_0\right>$ of a Hilbert space ${\Bbb H}$, with $T$ being a
representation of any Lie group operating on this ${\Bbb H}$.
Thus, applying this to our case, we can choose any spin state
$\left|m\right> \in {\Bbb H}_s=\{\left|m\right>|\, -s\leq m \leq s\}$
as starting vector together with the Lie group $SU_2$, leading to the
generalized coherent spin states
\be
\left|\bOmega, m\right>=e^{-iS_z\phi}e^{-iS_y\theta}e^{-iS_z\chi}
\left|m\right>,
\label{CS}
\ee
where $\phi$, $\theta$, and $\chi$ are the Euler angles.
Note that the dispersion of $\bS^2$ is not minimal for
$\left|m\right>\ne\left|s\right>$. However, as long as $1\ll |m|\lesssim
s$, the
spin fluctuations remain small. By means of Schur's Lemma, one can prove
as in Ref.~\onlinecite{Perelemov} that the integral
$\int d\Omega\;\left|\bOmega, m\right>\left<\bOmega, m\right|\propto
\cI$. So we have to calculate one single diagonal element in order to
fix
the proportionality constant. For this it turns out to be
convenient to express the states
$\left|\bOmega, m\right>$ by $2s$ spin-1/2 states that are individually
rotated, i.e.,
$\left|\uparrow'\right>=\cos(\theta/2)\left|\uparrow\right>
+e^{i\phi}\sin(\theta/2)\left|\downarrow\right>$,
$\left|\downarrow'\right>=e^{-i\phi}\sin(\theta/2)\left|
\uparrow\right>
-\cos(\theta/2)\left|\downarrow\right>$,
where the single-valued states are represented in the north pole gauge.
\cite{northpole}
Thus, we obtain
\be
\left|\bOmega, m\right>=\left(2s \atop s-m\right)^{-1/2}
\hspace{-0.5cm}
\sum_{j_1<\cdots<j_{s-m} \atop 1\le j_k\le 2s}
\left|\uparrow_1'\cdots\downarrow_{\{j_k\}}'\cdots\uparrow_{2s}'\right>,
\ee
where $\downarrow_{\{j_k\}}'$ denotes the $s-m$ rotated down spins
$\left|\downarrow_{j_k}'\right>$, at the positions $j_k$. Evaluation of
the amplitude
\bea
\left<\bOmega, m|s\right> & = &
\left<\bOmega, m|\uparrow_1\uparrow_2\cdots\uparrow_{2s}\right> \nn\\
& = & \sqrt{\left(2s \atop
s-m\right)}\left(\cos\frac{\theta}{2}\right)^{s+m}\left(e^{i\phi}\sin
\frac{\theta}{2}\right)^{s-m}
\label{amplitude_omega}
\eea
leads to
$\int d\Omega\;\left<s|\bOmega, m\right>\left<\bOmega, m|s\right>
={4\pi}/{(2s+1)}$.
Hence, the integration measure for all $\left|\bOmega, m\right>$ states is
the same.
Next, to derive the path integral, we need the quantity
$
\left<\tilde{\bOmega}, m\left|\bS\right|\bOmega, m\right>=
\left(m\bOmega+\cO(\sqrt{s})\right)
\left<\tilde{\bOmega}, m|\bOmega, m\right>
$,
\cite{fluctuations}
where we must choose the value of $m$ close to $s$.
Consequently, we have to evaluate the overlap between two states
separated by the angles $\delta\theta=\tilde{\theta}-\theta$ and
$\delta\phi=\tilde{\phi}-\phi$,
\bea
\left<\tilde{\bOmega}, m|\bOmega, m\right>
& = & \left(\text{terms containing
}\left<\uparrow'|\downarrow'\right>\text{ and }
\left<\downarrow'|\uparrow'\right>\right) \nn\\
& & +\left(\cos\frac{\tilde{\theta}}{2}\cos\frac{\theta}{2}
+\sin\frac{\tilde{\theta}}{2}\sin\frac{\theta}{2}e^
{-i(\tilde{\phi}-\phi)}\right)^{s+m}
\nn\\
& &
\times\left(\sin\frac{\tilde{\theta}}{2}\sin\frac{\theta}{2}e^{i(\tilde{\phi
}-\phi)}
+\cos\frac{\tilde{\theta}}{2}\cos\frac{\theta}{2}\right)^{s-m}. \nn
\eea
First-order expansion in $\delta\theta$ and $\delta\phi$
leads to the overlap between two
infinitesimally separated states,
$\left<\tilde{\bOmega}, m|\bOmega, m\right>\approx
1+im(\cos\theta-1)\delta\phi$.
After slicing the imaginary time interval $\beta$ into $N$ pieces
of length $\epsilon=\beta/N$,\cite{Fradkin,Loss} the transition amplitude
$\left<\bOmega, m(\tau_{n+1})\left|1-\epsilon\cH\right|\bOmega, m(\tau_n)\right>$
between two time steps $\tau_{n+1}$ and $\tau_n=n\epsilon$ can be
approximated by
$(1-\epsilon\cH[m\bOmega])\left<\bOmega, m(\tau_{n+1})|\bOmega, m(\tau_n)\right>$,
where $\cH[m\bOmega]$ is the diagonal element of the Hamiltonian.
Collecting the various terms, we obtain
\be
\left<\tilde{\bOmega}, m\left|e^{-\beta\cH}\right|\bOmega, m\right>=
\int\cD\Omega\; e^{-i\Phi(m)-S_{\rm dyn}},
\label{path_integral}
\ee
where $S_{\rm dyn}=\int_0^\beta d\tau\;\cH[m\bOmega]$ is
the dynamical action and $\Phi(m)=m\int_{-\pi/2}^{\pi/2} d\phi(1-\cos\theta)$ the
Berry phase, with $\Phi(m)/m$
measuring the area of the spherical triangle between the tunneling path
and the great circle through the north pole of $S^2$ (see
Fig.~\ref{interference2}).

Now we possess the tool to compute the  path integral
in Eq.~(\ref{amplitude}) also for low-lying excited states
$\left|m\right>$,
$m\lesssim s$. Since substitution of $s$ by $m$ in Eq.~(\ref{energy_y})
does not change the symmetry of $E_{y,n}$, $S_{\rm dyn}$ remains
invariant under
$C_{y,n}$ operations, whereas $\cR_{y,\eta}\Phi(m)/m=\Phi(m)/m+4\pi/n$. Thus, we can
divide the $S^2$ surface area of integration in
Eq.~(\ref{path_integral}) into $n$
%simply connected and
equally shaped
subareas $\cA_n$, so that we can factor out the following sum over Berry
phase terms without the need of evaluating the dynamical part of the
path integral (i.e., the following result is valid for all $m$),
\be
\left<-\frac{\pi}{2}\left|e^{-\beta\cH_{y,n}}\right|+\frac{\pi}{2}\right>
\propto
\sum_{k=1}^{n}e^{i(2\pi/n)(2k-1)m}
= \frac{\sin(2\pi m)}{\sin[(2\pi/n)m]},
\label{phases}
\ee
which vanishes whenever $n$ is not a divisor of $2m$. Thus, the tunnel
splitting energy $E_{m,-m}$ between the states $\left|m\right>$ and
$\left|-m\right>$ vanishes if $2m/n\not\in{\Bbb Z}$. However, if
$2m/n\in{\Bbb Z}$, the variable $m$ must be extended to real numbers,
i.e., $m\rightarrow \mu\in{\Bbb R}$, in order to calculate the limit
$\mu\rightarrow m$ of the ratio of the sine functions, which is plotted
in Fig.~\ref{phase} for a special case. In order to
visualize the interference between the Berry phases in
Eq.~(\ref{phases}), we select one representative path of each subarea
$\cA_n$. Then the vanishing of the amplitude in Eq.~(\ref{phases}) for
the
case $n=6$ (see Fig.~\ref{interference2}) can be thought of as a
destructive interference between six different paths. Note that there is
no simple pairwise cancellation.\cite{Delft}
Also,
it is important to note that from the semiclassical point of view the
total classical energy of the spin system $E$ must be conserved during
the
tunneling process. Therefore the semiclassical paths do not follow the
local minima of $E$ with respect to $\theta$ alone, except for the case
of
purely quadratic (i.e., $n=2$) anisotropies.\cite{Loss,Garg}

\begin{figure}[thb]
  \begin{center}
    \leavevmode
\epsfxsize=7cm
\epsffile{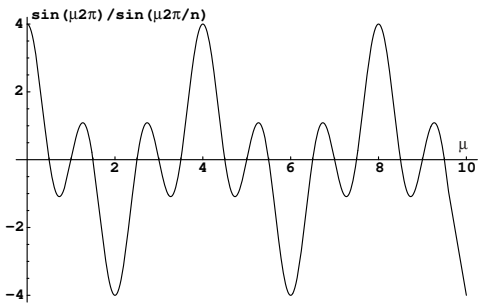}
  \end{center}
\caption{Sum over Berry phase terms for $s=10$ and $n=4$.}
\label{phase}
\end{figure}

{}From the above treatment we conclude that tunneling between two
degenerate spin states $\left|m\right>$ and
$\left|m'\right>=\left|-m\right>$ is topologically suppressed (i.e.,
the twofold degeneracy is not lifted) whenever
$n$ is {\it not} a divisor of $2m$. Since $n$ is even this excludes
immediately tunneling for all half-odd integer spins $s$ (for all $m$ and
$n$), in accordance with Kramers degeneracy. For the $s$ integer, however,
tunneling can be either allowed or suppressed,
depending on the ratio $2m/n$. In the latter case, the
twofold degeneracy of spin states is not lifted by the anisotropy, and
we
can view this result as a generalization of the Kramers theorem to
integer
spins.
For the special case $n=2$, we immediately recover the spin-parity effect
found before.\cite{Loss}

Now let us take a look at the tunnel splitting
energy
$E_{mm'}$ by means of higher-order degenerate perturbation theory in
terms
of a standard spin operator formalism.
Applying the resolvent technique\cite{LLMn12} we find
\be
E_{mm'}=2\left|
\frac{B_n^{(m-m')/n}}{\prod_{k=1}^{[(m-m')/n]-1}A[m^2-(m-kn)^2]}
\right|\delta_{m-m',jn},
\label{splitting}
\ee
where $j\in{\Bbb Z}$. Note that
$E_{mm'}$ vanishes if $2m/n\not\in{\Bbb Z}$ for $m'=-m$. This is a
consequence of the fact that $\cH_{y,n}$ divides the Hilbert space
${\Bbb H}_s$ into a direct sum of invariant
subspaces
$h_k=\{\left|m'\right>|\, s-m'=k\; {\rm mod} \, n\}$, i.e.,
${\Bbb H}_s=\bigoplus_{k=1}^{n}h_k$. Thus, the topological
interference effects due to
the Berry phase presented above correspond to these quantum-mechanical
selection rules.

Next we take the external transverse magnetic field $H_z$
for the case $n=2$ into account. For this, we consider the following
Hamiltonian ($A'=A-2B_2$, $B'=4B_2$):
\be
\cH_{y,2}=-A'S_y^2+B'S_z^2+h_zS_z+A'm^2C,
\label{H_y_hz}
\ee
yielding the classical energy
$E_{y,2}(u)=-A'm^2\left[\left(1-u^2\right)\sin^2\phi-C\right]+B'm^2u^2+h_zmu
$,
where $h_z=g\mu_BH_z$ and $u=\cos\theta$.
The constant $C=1$ is determined by the boundary condition\cite{Klauder}
$u(\phi=\pm\pi/2)=0$ for $h_z=0$.
By completing the square in $E_{y,2}(u)$ by the substitution
$\tilde{u}=u+h_z/2m(A'\sin^2\phi+B')$, the Jacobian of which is unity,
we end up with the anisotropy energy
\bea
E_{y,2}(\tilde{u}) & = &
-A'm^2\left[\left(1-\tilde{u}^2\right)\sin^2\phi-1\right]+B'm^2\tilde{u}^2
\nn\\
& & -\frac{1}{4}\frac{h_z^2}{A'\sin^2\phi+B'}.
\eea
$E_{y,2}(\tilde{u})$ and the new integration limits for $\tilde{u}$ are
invariant under $C_{y,2}$ operations. Therefore, we can factor out two
exponential terms with the following Berry phases:
\bea
\Phi_\pm(m) & = & m\int_{-\pi/2}^{\pi/2,-3\pi/2}d\phi
\left(1+\frac{h_z}{2m(A'\sin^2\phi+B')}\right) \nn\\
& = & \pm\pi m\left(1+\frac{h_z}{2m\sqrt{B'(A'+B')}}\right),
\eea
from which we directly obtain
$\left<-\pi/2\left|e^{-\beta\cH_{y,2}}\right|+\pi/2\right>
\propto \cos(\Phi_+)$  and thus, Berry phase oscillations in $H_z$ with
period
$2\sqrt{B'(A'+B')}=2\sqrt{4B_2(A+2B_2)}$, and
with phase shifts $|\Phi_\pm (m)-\Phi_\pm (m-1)|=\pi$,
regardless between which states $\left|m\right>$ and $\left|-m\right>$
the tunneling takes place. Since $0\le\left|\Phi_+-\Phi_-\right|\le 4\pi
m$, there are $2m$ zeroes of the tunnel splitting
$E_{m,-m}$.\cite{instanton} These
findings agree with the exact diagonalization of Eq.~(\ref{H_y_hz})
and have been observed in
Fe$_8$.\cite{Wernsdorfer}
Note that these interference effects are due to the symmetry $C_{y,2}$.
Hence, when breaking this symmetry with a magnetic field that is tilted away
from
the $z$ axis (in the $xz$ plane) by the angle $\psi$,
the oscillations become  suppressed with
increasing $\psi$,\cite{negative_B} in agreement with
Ref.~\onlinecite{Wernsdorfer}. In contrast to previous work on $n=2$
(Ref.~\onlinecite{Garg}) we did
not use approximations based on semiclassical paths (given by energy
conservation) or on semiclassical Wentzel-Kramers-Brillouin methods\cite{Garg2} to find the
zeroes of the tunnel splitting.

Next, we consider the case $n=4$,
\be
\cH_{y,4}=-AS_y^2+B_4\left(S_+^4+S_-^4\right)+h_zS_z+Am^2C,
\ee
and its corresponding classical energy
\bea
E_{y,4} & = & -Am^2\left[(1-u^2)\sin^2\phi-C\right]
+2B_4m^4\left[u^4-6u^2\right. \nn\\
& & \left.\times(1-u^2)\cos^2\phi+(1-u^2)^2\cos^4\phi\right]+h_zmu.
\label{quartic}
\eea
Again, the boundary conditions determine the constant $C=1$ for $h_z=0$.
Unfortunately, it is no longer possible to complete the quartic
polynomial in Eq.~(\ref{quartic}). The same problem persists for $n>4$.
This means that in contrast to $n=2$, the $C_{y,4}$ symmetry of the
anisotropy energy $E_{y,4}$ without transverse field $H_z$ cannot be
restored anymore. However, there still remains the reflection symmetry
between the conjugated paths $\gamma_+=(\theta,\phi)$ and
$\gamma_-=(\theta,-\phi)$ that wind around the $z$ axis counterclockwise
and clockwise, respectively. The topological part of the main
contribution to the tunneling amplitude exhibits oscillations, as shown
in
Fig.~\ref{berryphase}. This part comes from the two
semiclassical paths $u_{{\rm cl},\pm}(\theta)$ that satisfy $E_{y,4}=0$
and whose Berry phases, defined by
\be
\pm\Phi_{\rm cl}=m\int_{-\pi/2}^{\pi/2,-3\pi/2}
d\phi(1-{\rm Re}\{u_{{\rm cl},\pm}\})\, ,
\label{berryclassical}
\ee
interfere
with each other.  Thus, the level splitting induced by tunneling
shows quantum
oscillations as a function of $H_z$ due to this oscillating Berry
phase.\cite{instanton}
Again, for a magnetic field with $\psi\ne k\pi/2$, $k\in\Bbb{Z}$,
the reflection symmetry between conjugated paths is broken (see above),
and therefore the more $\psi$ is detuned from $k\pi/2$ the more the
Berry phase oscillations are suppressed.\cite{negative_B}
Note that the level splitting contains $m$ zeroes, $m=8,9,10$,
in the interval $-4$ T$\le H_z\le 4$ T
(see Fig.~\ref{berryphase}), and that the Berry phase
oscillations are shifted by $\pi/2$  when the parity of $m$ changes
(see $m=8,9,10$ in Fig.~\ref{berryphase}), all in agreement with exact
diagonalization.\cite{Tupitsyn}  Thus, we have obtained a new spin-parity effect for
$n=4$ in the case of integer spins $s$ which results
from Berry phase oscillations. This effect should be observable for
$\psi=k\pi/2$ in
molecular magnets such as Mn$_{12}$.
Since for all $n$ two conjugated paths (see above) can be found, Berry
phase oscillations and their associated parity effect are maximal
whenever $\psi=k2\pi/n$.\cite{negative_B} This provides an experimental means to determine
the order $n$ of
transverse  anisotropies.

\begin{figure}[thb]
  \begin{center}
    \leavevmode
\epsfxsize=8cm
\epsffile{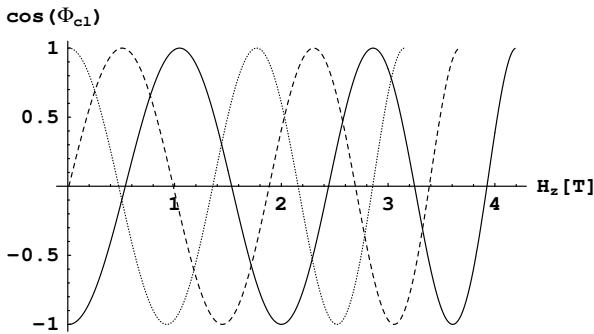}
  \end{center}
\caption{Topological part of the tunneling amplitude for $n=4$ and
$s=10$ as a function of $H_z$ [see Eq.~(\protect\ref{berryclassical})]. Solid
line:
$m=10$. Dashed line: $m=9$. Dotted line: $m=8$. The right boundary of
$\cos(\Phi_{\rm cl})$ is determined by $0\le 2|\Phi_{\rm cl}|\le 4\pi
m$.
The values $A=0.54$ K and $B_4=8.5\times10^{-5}$
K approximate roughly the anisotropies in
Mn$_{12}$ (Ref.~\protect\onlinecite{LLMn12})}
\label{berryphase}
\end{figure}

In conclusion, we have shown that
 the tunnel splitting vanishes
due to interfering spin Berry phase and thus the level degeneracy is not
lifted
whenever $n$ is not a divisor of $2m$. This generalization of the
Kramers degeneracy
to integer spins
holds not only for the ground state
but also for low-lying excited states $m$
and for transverse
anisotropies of arbitrary order $n$; the presence
of magnetic fields leads to oscillations of the tunnel splitting.

This work has been supported in part by the Swiss NSF and by Molnanomag HPRN-CT-1999-00012.


\begin{references}
\bibitem{northpole}
We note that the chosen parametrization makes the spin-1/2 states
single-valued everywhere on the sphere except at the south pole (i.e.,
the Dirac string is placed along the negative $z$ axis) (see Ref.~\protect\onlinecite{Loss}).
\bibitem{fluctuations} $\left<\tilde{\bOmega},m|\bOmega,m\right>$ has a width of order $\sqrt{s}$ if $m$ is close to $s$.
\bibitem{instanton} It is well known from instanton
calculations (Ref.~\onlinecite{Loss})
that
$E_{m,-m}\propto\left<-\pi/2\left|e^{-\beta\cH_{y,2}}\right|
+\pi/2\right>$
and thus $\propto|\cos(\Phi_+)|$.
\bibitem{negative_B} Changing the sign of $B_n$ is equivalent to applying $\cR_{\alpha,\eta/2}$ to $E_{\alpha,n}$. Thus for $B_n<0$ the Berry phase oscillations are maximal for $\psi=j\pi/n$,
where $j$ is an odd integer.
\end{references}
\end{document}